\documentclass[
reprint,
amsmath,amssymb,
aps,prx,
floatfix,
longbibliography
]{revtex4-2}

\usepackage{graphicx}
\usepackage{dcolumn}
\usepackage{bm}
\usepackage{xcolor}
\usepackage{amsmath,amssymb}
\usepackage{hyperref}
\hypersetup{
	colorlinks=false,  
	pdfborder={0 0 0}, 
	breaklinks=true
}

\begin{document}

\title{Subcritical transition to turbulence in buoyancy-driven flows with multiple hysteresis loops under quasi-one-dimensional confinement}

\author{Lu Zhang}
\email{zhangl39@sustech.edu.cn}
\affiliation{Center for Complex Flows and Soft Matter Research, Department of Mechanics and Aerospace Engineering, Southern University of Science and Technology, Shenzhen 518055, China}

\author{Ke-Qing Xia}
\email{xiakq@sustech.edu.cn}
\affiliation{Center for Complex Flows and Soft Matter Research, Department of Mechanics and Aerospace Engineering, and Department of Physics, Southern University of Science and Technology, Shenzhen 518055, China}

\date{\today}

\begin{abstract}

We present both static and quasi-static direct numerical simulations of Rayleigh--B\'enard convection in a quasi-one-dimensional domain, revealing for the first time a clear subcritical transition to turbulence in a buoyancy-driven flow. Within a narrow range of Rayleigh number ($Ra$), three coexisting flow states are identified: steady convection, oscillatory chaos, and intermittent turbulence. The transitions between these states are accompanied by abrupt jumps in both the Nusselt number ($Nu$) and Reynolds number ($Re$), the key global transport quantities in buoyancy-driven flows. Additionally, they exhibit pronounced hysteresis, forming three distinct hysteresis loops in the $Nu$-$Ra$ plane: normal, reverse, and anomalous. More importantly, we show that the steady convection state is linearly stable against infinitesimal perturbations but can transition to intermittent turbulence when subjected to finite-amplitude disturbances, which is a defining hallmark of subcriticality. Thus, contrary to the prevailing view that the transition from convection to turbulence is supercritical, our results demonstrate that buoyancy-driven turbulence can emerge via a subcritical route, paving the way for a unified framework that describes instability mechanisms in both buoyancy-driven and shear-driven flows.

\end{abstract}

\maketitle

\section{Introduction}

The transition from laminar to turbulent flow remains a core issue in the study of fluid turbulence, which governs the transport of momentum and energy in almost all natural and engineering flow systems \cite{Drazin_Book_1981,Pope_Book_2000}. It has attracted researchers across physics, mathematics, and engineering over the past several decades, leading to the identification of two fundamentally distinct transition pathways: subcritical and supercritical transitions. Subcritical laminar-to-turbulent transition is ubiquitous in canonical shear flows, such as pipe and channel flows \cite{Reynolds_PhilTrans_1883,Pomeau_PhysicaD_1986,Trefethen_Science_1993,Reddy_JFM_1998,Avila_Science_2011}. It is characterized by an abrupt jump in the friction coefficient, coexistence of laminar and turbulent states, and pronounced hysteresis \cite{Moody_TASME_1944}. In contrast, supercritical transition is commonly associated with buoyancy-driven flows such as classical Rayleigh--B\'enard convection (RBC), proceeding through a sequence of bifurcations without hysteresis or coexisting states \cite{Chandrasekhar_Book_1961,Krishnamurti_JFM_1970,Clever_JFM_1974,Ahlers_PRL_1978,Gollub_JFM_1980}. 
Another essential difference is that a subcritical transition occurs in linearly stable flows that requires finite-amplitude disturbances to induce turbulence, whereas a supercritical transition in buoyancy-driven flow arises spontaneously through linear instability when the driving force (usually quantified by the Rayleigh number $Ra$) exceeds a critical value.

Recently, subcritical laminar-to-turbulent transition in shear-driven flow has seen remarkable advances, with an analogy to the directed percolation universality class in statistical mechanics \cite{Hof_2022_NP,Lemoult_2024_NP}. By contrast, theoretical understanding of buoyancy‑driven flows is largely limited to the supercritical transition from conduction to steady convection (laminar flow) \cite{Chandrasekhar_Book_1961}, and the secondary instability leading to periodic oscillation \cite{Busse_RPP_1978}. This raises the following fundamental question: Does any alternative pathway to turbulence, such as subcritical transition, exist in buoyancy-driven flows? If this is the case, the two seemingly distinct classes of basic flows--buoyancy-driven flow and shear-driven flow--can be understood within a single universal theoretical framework.


Thermal convection presents one of the simplest examples of hydrodynamic instability, and serves as an idealized platform for studying buoyancy-driven flow. Celebrated works on thermal convection \cite{Krishnamurti_JFM_1970,Clever_JFM_1974,Ahlers_PRL_1978,Libchaber_JPL_1978} reveal that, due to the coupled advections of both momentum and temperature, the primary transition from conduction to steady convection and the subsequent transition from steady flow to periodic oscillatory instability are both supercritical. The first is associated with the Rayleigh--B\'enard instability, where a transition from conduction to steady convection at a critical Rayleigh number $Ra_c = 1708$ for an infinite plate is predicted theoretically and observed experimentally \cite{Chandrasekhar_Book_1961}. With a further increase in $Ra$, the onset of oscillatory instability is manifested by a transition from time-independent convection to periodic oscillation. Subsequent bifurcations exhibit more complex time dependence, including quasi-periodicity, chaos, and finally, fully developed turbulence. In addition, throughout the above supercritical transitional processes, the global heat and momentum transport of the system, quantified by the Nusselt number $Nu$ and Reynolds number $Re$ respectively, both vary continuously with the thermal driving strength $Ra$.

Over the past few decades, many progresses have been achieved on thermal convection research at high Rayleigh number turbulent state, with major topics including heat transport, flow dynamics, boundary layer properties, and small-scale turbulence \cite{Siggia_ARFM_1994,Niemela_Nature_2000,Grossmann_JFM_2000,Ahlers_RMP_2009,Lohse_ARFM_2010,Chilla_EPJE_2012,Xia_TAML_2013}. Notably, even when turbulent convection is fully developed in the bulk region, the boundary layers can still remain laminar. Recent studies reveal that, at sufficiently high $Ra$, the Prandtl--Blasius-type laminar boundary layer in thermal convection undergoes a nonnormal and nonlinear subcritical transition toward turbulent boundary layer \cite{Shishkina_PRL_2024,Lohse_RMP_2024}. Furthermore, under external forcing or external fields, subcritical onset of convection has also been reported \cite{Ding_JFM_2022,McCormack_JFM_2025}.

On the other hand, although most existing studies on thermal convection consider convection domains with aspect ratio around unity or larger, recent studies on severely confined quasi-two-dimensional and quasi-1D RBC demonstrate that lateral confinement strongly affects global transport, \cite{Huang_PRL_2013, Chong_PRL_2015, Ahlers_PRL_2022, Zhang_JFM_2023,Xia_NSR_2023}, thereby suggesting a possible influence of dimensionality on the transition from convection to turbulence \cite{Ren_JFM_2024,Zheng_JFM_2025}.

Dimensionality governs the intrinsic symmetry and degrees of freedom of a physical system. Therefore, reducing the dimensionality often leads to unprecedented changes in macroscopic properties and emergence of novel phenomena. Examples abound in statistical mechanics, quantum mechanics, and fluid dynamics. For instance, the classical two-dimensional Ising model undergoes a sharp finite-temperature phase transition \cite{Onsager_PR_1944}, whereas no such transition exists in one-dimensional Ising model; the two-dimensional electron gas in a strong magnetic field displays the quantum Hall effect \cite{Klitzing_PRL_1980}, which is suppressed in three-dimensional systems; and the  phenomenon of inverse energy cascade in two-dimensional turbulence \cite{Boffetta_ARFM_2012} is also a direct consequence of reduced dimensionality.

The present work aims to uncover the role of dimensionality on the laminar-to-turbulent transition in buoyancy-driven flows. Specifically, we focus on the transition of flow states in a highly confined quasi-1D RBC. Combining the conventional static simulation and a novel quasi-static numerical protocol introduced in this work, we identify three coexisting flow states along the transition pathway: steady convection, oscillatory chaos, and intermittent turbulence. Remarkably, the transition pathway in quasi-1D RBC bypasses the single-frequency periodic oscillatory state that is regarded as a key intermediate state in laterally unconfined convection systems. In addition, the route from steady convection to intermittent turbulence exhibits three distinct hysteresis loops. Finally, stability analysis provides definitive evidence that the laminar-to-turbulent transition in quasi-1D RBC is subcritical.

\section{Methods\label{Methods}}
We consider a rectangular quasi-1D RBC system with a square horizontal cross-section and a lateral-to-vertical aspect ratio of $\Gamma = 0.1$. The Prandtl number is fixed at $Pr = 4.34$ (corresponds to water at 40 $^\circ$C ). The governing equations are solved using the $CUPS$ code \cite{Chong_JCP_2018}, in the following non-dimensional form: 

\begin{equation}
	\frac{\partial \textbf{\textit{u}}}{\partial t} + \textbf{\textit{u}}\cdot\nabla\textbf{\textit{u}} = -\nabla p + \sqrt{{Pr}/{Ra(t)}}\nabla^2\textbf{\textit{u}} + T\hat{\textbf{\textit{z}}}, 
\label{nondim_mom}
\end{equation}

\begin{equation}
	\frac{\partial T}{\partial t} + \textbf{\textit{u}}\cdot\nabla T = 1/\sqrt{{PrRa(t)}}\nabla^2T,
\label{nondim_temp}
\end{equation}

\begin{equation}
	\nabla\cdot \textbf{\textit{u}} = 0.
\label{nondim_incomp}
\end{equation}

In the quasi-static numerical scheme, Rayleigh number is allowed to vary continuously with time $Ra(t) = Ra_0 + ct$. Here, $Ra_0$ is the initial Rayleigh number, and $c$ denotes the temporal variation rate of Rayleigh number. Practically, we update $Ra(t)$ according to this formula at every time step, which is fixed at $10^{-3}$ free-fall time units in this study. With a sufficiently small $c$, such a simulation protocol is analogous to the quasi-static process in thermodynamics.

In contrast, conventional static simulations adopt fixed discrete Rayleigh numbers, in which the flow field evolves directly from an initial state at $Ra_0$ to a target Rayleigh number of $Ra_1 = Ra_0+\Delta Ra$. Compared with static simulations, the quasi-static numerical scheme enables both efficient exploration of parameter space and reliable identification of potential hysteresis behaviors.

Both \textit{static} and \textit{quasi-static} simulations are performed over a Rayleigh number range \(7.9\times10^7\le Ra\le8.5\times10^7\). For the aspect ratio \(\Gamma=0.1\) quasi-1D rectangular cell considered in this study, the onset Rayleigh number is approximately \(Ra_c = 7.2\times10^6\) \cite{Shishkina_JFM_2021,Zhang_JFM_2023,Ren_JFM_2024}. Accordingly, the reduced Rayleigh number falls within a narrow range of \(11.0\le Ra/Ra_c\le11.8\). For static simulations, the Rayleigh number increment is fixed to be $|\Delta Ra|=2\times10^5$. While for quasi‑static simulations, the absolute variation rate is set as $|c|=20$ or $50$. This guarantees that change of $Ra(t)$ is sufficiently slow compared to the typical time scale of the convective flows, thus satisfying the quasi‑static approximation. 

\begin{figure*}[htbp!]
	\centerline{\includegraphics[width=2.0\columnwidth]{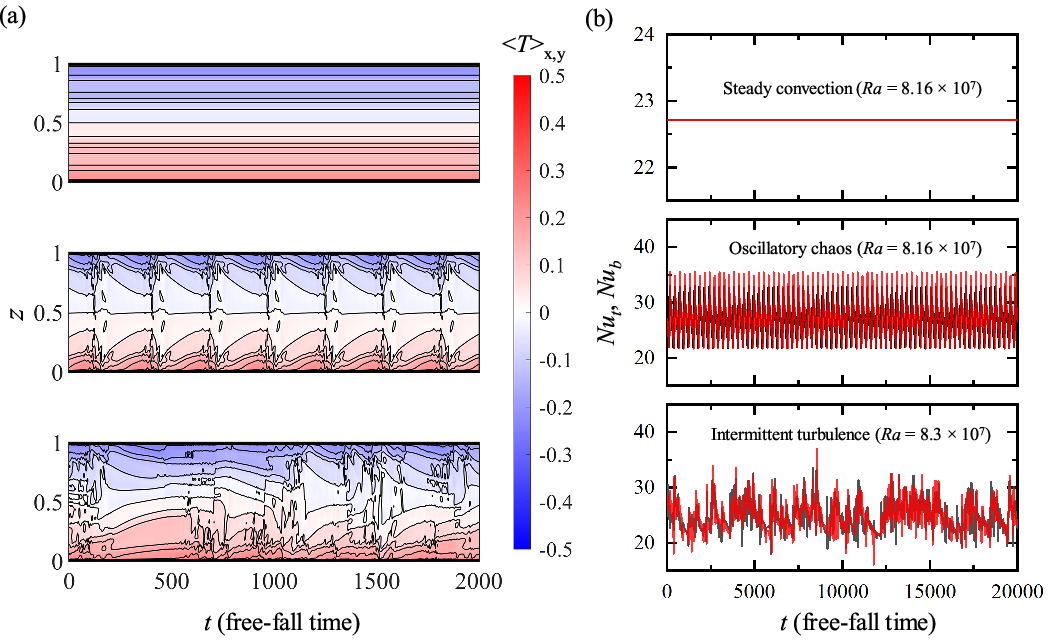}}
	\caption{(a) Spatiotemporal diagrams showing the evolution of horizontally averaged temperature profiles for three distinct flow states: steady convection, oscillatory chaos, and intermittent turbulence. Red and blue denote hot and cold regions, and solid lines represent different isothermals. (b) Corresponding time series of the top- and bottom-plate Nusselt numbers $Nu_t$ (red) and $Nu_b$ (black). For the convection state, the two are identical.}
	\label{fig:threestates}
\end{figure*}

\section{Results and discussions\label{Results}}

\subsection{Steady convection, Oscillatory chaos and Intermittent turbulence\label{ThreeStates}}

In this subsection, we present the results obtained by conventional \textit{static} simulation. Figure~\ref{fig:threestates} shows the three distinct flow states observed throughout the transition in the quasi-1D RBC. The space-time diagrams of the horizontally averaged temperature profile $\langle T\rangle_{x,y}$ [Fig.~\ref{fig:threestates}(a)] clearly illustrate their intrinsic differences: The top panel corresponds to steady convection, where the temperature profile remains time-independent with no discernible temporal variation. The middle panel represents oscillatory chaos, characterized by a quasi-periodic oscillation, with a vertebra-shaped spatiotemporal pattern. The flow alternates between quiescent relaxation stages and intense convective bursts, reflecting slow thermal accumulation near the boundaries followed by rapid, eruptive heat release. The lower panel shows intermittent turbulence, dominated by irregular slow variations and abrupt turbulent bursts without evident periodicity. 
These distinctions are also reflected in the global heat transport shown in Fig.~\ref{fig:threestates}(b). The red and black lines represent the time series of Nusselt number measured at the top plate $Nu_t = -\langle\partial T/\partial z\rangle_{z=H}$ and at the bottom plate $Nu_b = -\langle\partial T/\partial z\rangle_{z=0}$, respectively. For steady convection, $Nu_t$ and $Nu_b$ both remain constant over time, consistent with time-invariant laminar flow. In the oscillatory chaotic state, both Nusselt numbers exhibit large-amplitude, quasi-periodic oscillations. The two curves nearly overlap with each other, though a small discrepancy is present, suggesting a broken up-down symmetry of the flow. By contrast, the time series of the intermittent turbulent state loses periodicity and is characterized by intermittent bursts or clusters of bursts separated by relatively quiescent intervals---a typical signature of turbulence. 


\begin{figure}[htbp!]
  \centerline{\includegraphics[width=1.0\columnwidth]{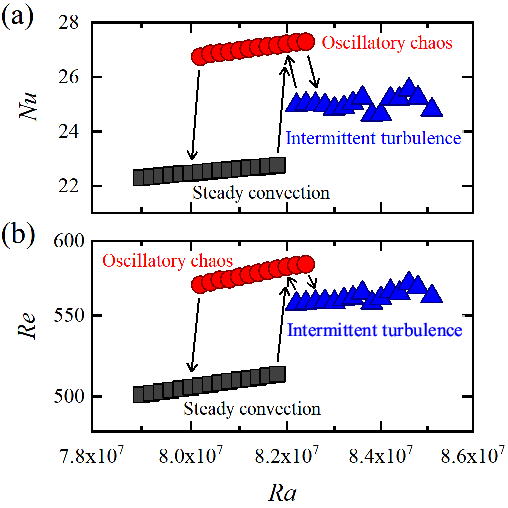}}
  \caption{Global transport properties of quasi-1D RBC with an aspect ratio of $\Gamma = 0.1$ obtained from static simulations. (a) The $Nu-Ra$ relationship. (b) The corresponding $Re-Ra$ relationship. The different symbols denote different flow states: steady convection (black squares), oscillatory chaos (red circles), and intermittent turbulence (blue triangles). The arrows indicate the hysteretic transitions between different flow states (see Fig.~\ref{fig:transitions} for details).
}
\label{fig:NuReRa}
\end{figure}

Figures \ref{fig:NuReRa}(a) and \ref{fig:NuReRa}(b) present the global $Nu-Ra$ and $Re-Ra$ relations obtained from \textit{static} simulations. Here, the mean Nusselt number is defined as $Nu = -(\langle
\partial T/\partial z\rangle_{z=0} +\langle \partial T/\partial z\rangle_{z=H})/2$, and the Reynolds number is defined based on the volume-averaged root-mean-square velocity: $Re = (Ra/Pr)^{1/2}\sqrt{\langle\mathbf{u}\cdot\mathbf{u}\rangle}_V$. Within the narrow Rayleigh number range of $8.02\times10^7\le Ra\le8.24\times10^7$, the heat and momentum transport properties are not single-valued functions of $Ra$. Instead, three statistical steady states coexists: an upper branch with the largest $Nu$ and $Re$ corresponding to oscillatory chaos, a lower branch with the lowest transport efficiencies corresponding to steady convection, and an intermediate branch corresponding to intermittent turbulence. In particular, the upper and lower branches coexist over $8.02\times10^7\le Ra\le8.18\times10^7$, with a 19\% deviation in $Nu$ and a 13\% deviation in $Re$, respectively. The upper and intermediate branches coexist within $8.22\times10^7\le Ra\le 8.24\times10^7$, where the oscillatory chaotic state exceeds the intermittent turbulent state by 9\% in $Nu$ and 3.5\% in $Re$. For the Rayleigh number range explored, oscillatory chaos is most efficient in heat transfer and most energetic in flow strength, while steady convection exhibits the lowest transport efficiencies of both heat and momentum. For convenience, these three states are hereinafter referred to as convection, chaos and turbulence, respectively. 

\begin{figure}[htbp!]
  \centerline{\includegraphics[width=1\columnwidth]{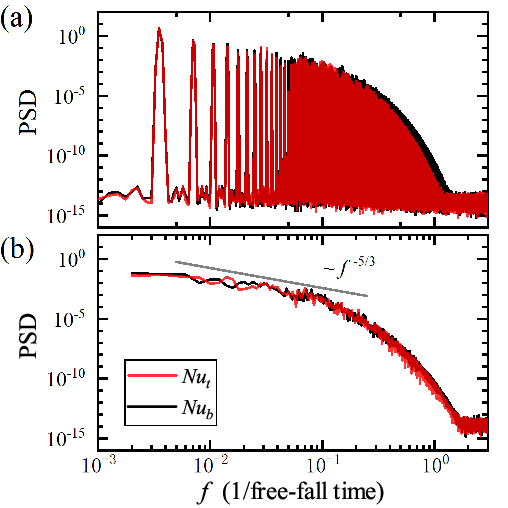}}
  \caption{Power spectral density (PSD) of the Nusselt number time series. (a) Oscillatory chaotic state. (b) Intermittent turbulent state, where the gray solid line indicates the Kolmogorov $-5/3$ scaling law. The red and black curves correspond to $Nu_t$ and $Nu_b$, respectively.}
  
\label{fig:PSD}
\end{figure}

The differences between chaos and turbulence are further elucidated in Fig.~\ref{fig:PSD} with power spectral density (PSD) analysis of their Nusselt number time series. In the oscillatory chaotic state [Fig.~\ref{fig:PSD} (a)], the spectra of $Nu_t$ and $Nu_b$ are dominated by a sharp fundamental peak at $f=3.56\times10^3$ (corresponding a period of $\tau=281$ free-fall time units), together with its multiple harmonics $2f$, $3f$, $4f$,$\dots$. Such a discrete spectrum confirms the quasi-periodic dynamic behavior of organized oscillatory motion. A small but persistent deviation between the two spectra at high frequencies also signals a broken up-down symmetry, consistent with our observations in Fig.~\ref{fig:threestates}.
In contrast, the intermittent state [Fig.~\ref{fig:PSD}(b)] shows a broadband decaying spectrum, and a region close to the classical Kolmogorov $-5/3$ scaling law is observed. The disappearance of discrete frequency peaks also confirms that the flow has transitioned from organized oscillation to irregular, stochastic dynamics, consistent with the intermittent turbulent behavior observed in the time-domain signals.

\begin{figure*}[ht!]
  \centering
  \includegraphics{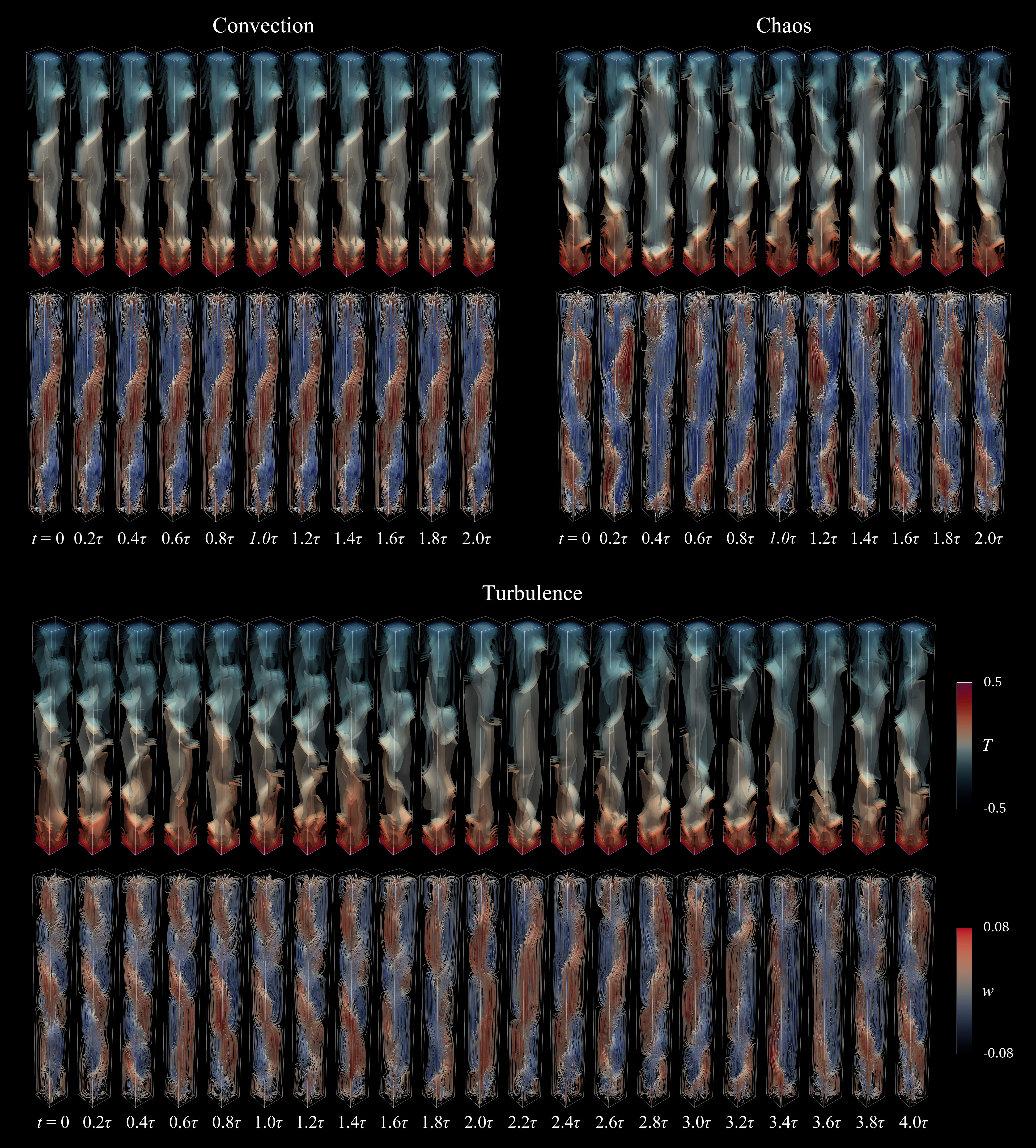}
  
  \caption{Snapshots of temperature isosurfaces (first and third rows) and streamlines (second and fourth rows) at three distinct flow states: steady convection (top left, $Ra=8.16\times10^7$), oscillatory chaos (top right, $Ra=8.16\times10^7$), and intermittent turbulence (bottom, $Ra=8.30\times10^7$).}

\label{fig:snapshots}
\end{figure*}

To further illustrate the flow structure of the three states, we plot in Fig.~\ref{fig:snapshots} the evolution of temperature field and the corresponding streamlines. For the steady convective state at $Ra=8.16\times10^7$ (top-left two panels), the flow maintains a stable, vertically stretched, torsional large-scale circulation \cite{Iyer_PNAS_2020, Zheng_JFM_2025}. 

At the same Rayleigh number, the oscillatory chaotic state (top-right two panels) retains the underlying elongated torsional circulation structure but undergoes clear quasi-periodic deformation. The temperature and velocity fields evolve with the dominant oscillation cycle, as evidenced by the repeating twisting motion of the plume structure over the $2.0\tau$ time span. We also note that in this oscillatory chaos state, the temperature and velocity fields undergo a horizontal mirror-flip every oscillation period \(\tau\), and the flow returns to its original structure after a duration of \(2\tau\). As a result, the main period of flow evolution is twice that of the dominant oscillation cycle obtained from the Nusselt number time series.
 
For the intermittent turbulent state at $Ra=8.30\times10^7$ (bottom two panels), even though the underlying circulation shows qualitative similarity with those in convection and chaotic state, the ascending and descending flows exhibit highly irregular dynamics, characterized by the random appearance, disappearance, and vertical drift of internal `twisting' or `crossing' events. These structural irregularities are consistent with recent experimental observations reported in \cite{Zheng_JFM_2025}.


\subsection{Transition between the three states\label{Transitions}}

\begin{figure}[htbp!]
  \centerline{\includegraphics[width=1.0\columnwidth]{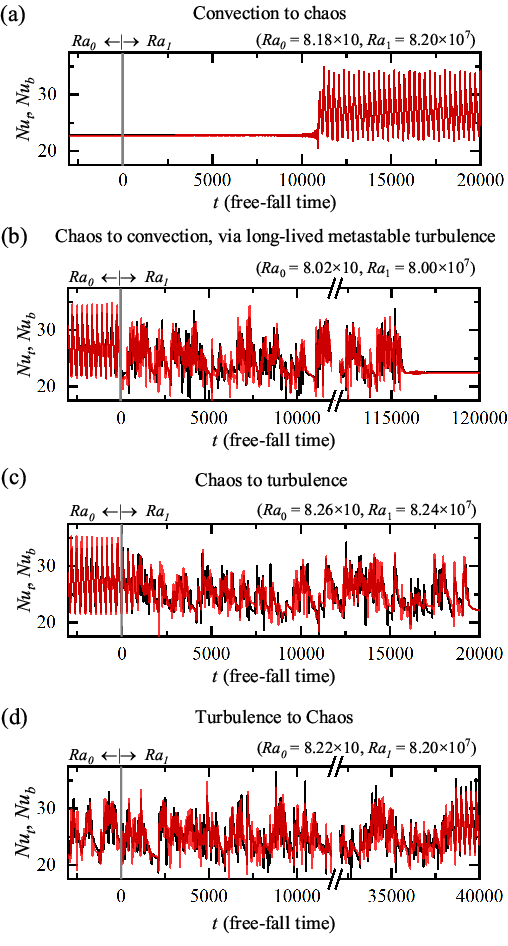}}
  \caption{Time series of top-plate (red) and bottom-plate (black) Nusselt numbers during transitions between convection and chaos (a-b), and those between chaos and turbulence (c-d) for a fixed Rayleigh number step of $|\Delta Ra| = 2\times10^5$. The vertical gray line denotes the time when the Rayleigh number were changed. An extremely long-lived (over $10^5$ free-fall time units) meta-stable intermittent turbulence is observed during the transition from chaos ($Ra_i = 8.02\times10^7$) to convection ($Ra = 8\times10^7$).}
\label{fig:transitions}
\end{figure}

The substantial, discontinuous gaps in both $Nu$ and $Re$ between the three coexisting states also indicate that the transitions between them are abrupt, in sharp contrast to the continuous, smooth variations observed in 3D RBC with aspect ratio around unity \cite{Heslot_PRA_1987,Castaing_JFM_1989,Niemela_Nature_2000}. Here, we examine these transitions using static simulations with a fixed Rayleigh number step of $\Delta Ra = 2 \times 10^5$. Even for such a small step, convergence to the new state can take up to $O(10^6)$ free-fall time units, as illustrated in the following. 

Figure \ref{fig:transitions} presents the time series of $Nu_t$ and $Nu_b$ during different transitions. It is noteworthy that the forward and reverse processes between each pair of states occur at different $Ra$ values, indicating strong hysteresis. First, the convection-to-chaos transition at $Ra_0 = 8.18\times10^7$ and $Ra_1 = 8.20\times10^7$ takes $O(10^4)$ time units [Fig.~\ref{fig:transitions}(a)]---longer than the chaos-to-turbulence transition. Most strikingly, its reverse, the chaos-to-convection transition, observed at $Ra_0=8.02\times10^7, Ra_1=8\times10^7$, proceeds via an extremely long-lived meta-stable intermittent turbulence state, persisting for over $O(10^6)$ free-fall time units [Fig.~\ref{fig:transitions}(b)]. This long-lived transient behavior is reminiscent of the growth of puff lifetimes near the laminar–turbulent transition in pipe flows \cite{Hof_2022_NP, Lemoult_2024_NP}. 

The transition from chaos to turbulence occurs at $Ra_0 = 8.02\times10^7$ and $Ra_1=8\times10^7$ [Fig.~\ref{fig:transitions}(c)], and such a transition process takes only about $O(10^3)$ free-fall time units, after which the oscillation suddenly loses periodicity and transitions into an intermittent turbulence state. In contrast, the reverse transition, from turbulence to chaos, occurs at $Ra_0=8.22\times10^7$ and $Ra_1=8.20\times10^7$ [Fig.~\ref{fig:transitions}(d)]. Compared to the transition from chaos to turbulence, the duration of the reverse transition is very long: the system remains in the intermittent turbulence state for over 38,000 time units before spontaneously switching to the oscillatory chaotic state. It is also worth mentioning that the time required for transitions between different flow states in quasi-1D RBC is orders of magnitude longer than the typical convergence times in 3D or 2D simulations \cite{Pandey_JFM_2025}. Furthermore, all transitions are irreversible and path-dependent, with forward and reverse processes occurring at distinct Rayleigh numbers. Taken together, the hysteresis behaviour and the abrupt jump in global heat transport during the transitions serve as hallmarks of subcritical transition. These key features, which are well-known in shear-driven flows such as pipe flow, have not been found in any buoyancy-driven flow systems.

\subsection{Three hysteresis loops\label{hysteresisLoops}}
The coexistence of multiple stable flow states is inherently linked to hysteresis, as suggested by our static simulations. To systematically explore the hysteretic behavior across the transition region, we employ the \textit{quasi-static} numerical protocol, which allows us to efficiently `scan' the relevant range of Rayleigh numbers at reduced computational cost.

Using this approach, we identify three distinct hysteresis loops. Two of these correspond to the chaos–turbulence and convection–chaos transitions already observed in our static simulations. The third, however, corresponds to a hysteresis loop between convection and turbulence---with transitions not yet captured by the discrete, step-wise static simulations. We begin our discussion with this newly identified loop.

\begin{figure}[htbp!]
  \centerline{\includegraphics[width=1.0\columnwidth]{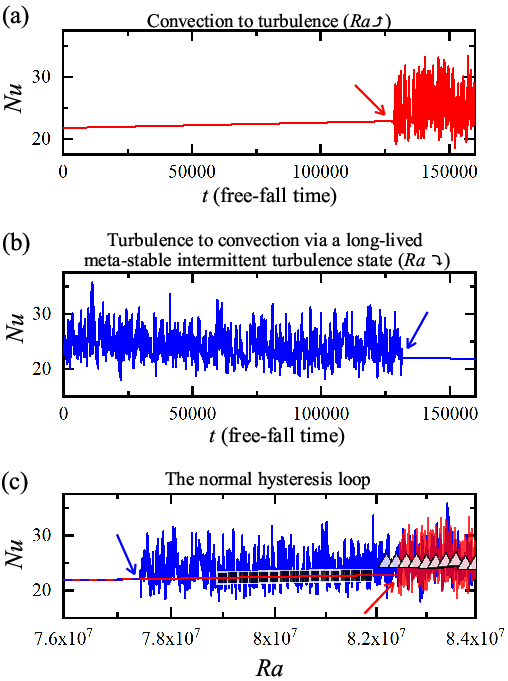}}
  \caption{(a) The time series of $Nu$ during a forward quasi-static process starting from $Ra_0 = 7.6\times10^7$ with a rate of $c = 50$. (b) The corresponding `inverse' process starting from $Ra_0 = 8.4\times10^7$ with $c = -50$. (c) The `normal' hysteresis loop between convection and turbulence in the $Nu-Ra$ plane. The solid squares and open triangles show data from static simulations. The arrows mark the locations where transitions occur. 
}
\label{fig:normal_loop}
\end{figure}

\subsubsection {The normal hysteresis loop}
Figure~\ref{fig:normal_loop} presents the quasi-static simulations that uncover the hysteretic transitions between convection and turbulence.

In the forward quasi-static process with $c=+50$, starting from a steady convective state at $Ra_0=7.6\times10^7$ [Fig.~\ref{fig:normal_loop}(a)], the Nusselt number first increases linearly with $Ra$. At $t\approx130,000$ free-fall time units, however, $Nu(t)$ suddenly exhibits large, irregular fluctuations, marking an abrupt transition from convection to turbulence at approximately $Ra = 8.25\times10^7$ (see the red arrow).

For the reverse process (Fig.~\ref{fig:normal_loop}(b)), which starts from the turbulence state at $Ra_0=8.4\times10^7$ with $c = -50$, the system remains in intermittent turbulence for over $130,000$ free-fall time units, before abruptly transitioning into a convective state, where all intense turbulent fluctuations (including $Nu$, temperature and velocity fields throughout the entire convection cell) suddenly come to a halt (see the blue arrow). Afterwards, the global heat transfer exhibits only a mild linear decrease over time.

The $Nu-Ra$ behavior of both processes is shown in panel Fig.~\ref{fig:normal_loop}(c), together with the static simulation data. The forward trajectory follows the lower branch of steady convection (orange line) until it bifurcates onto the intermittent turbulence branch at $Ra\approx 8.25\times10^7$. The reverse trajectory, in contrast, stays on the turbulence branch far beyond this point, before eventually jumping back to the convective branch at $Ra\approx 7.75\times10^7$, indicated by the arrow. Once the system returns to the convective state, its $Nu-Ra$ response collapses perfectly onto the forward-process curve.

Together, the forward and backward processes form a closed hysteresis loop in the $Nu-Ra$ plane between convection and turbulence, with two bifurcation points located near $Ra = 7.75\times10^7$ and $Ra=8.25\times10^7$. This hysteresis loop is analogous to the classical magnetization hysteresis loop observed in ferromagnetic materials, motivating our designation of it as the `normal' hysteresis loop.

\begin{figure}[htbp!]
  \centerline{\includegraphics[width=1\columnwidth]{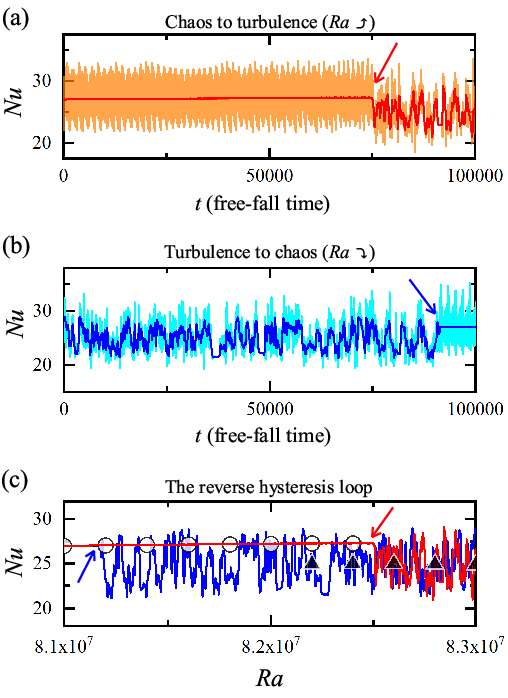}}
  \caption{(a) Time series of the Nusselt number (orange, raw data; red, running average) during a quasi-static simulation starting from $Ra_0 = 8.10\times10^7$ with a rate of $c = 20$. (b) Corresponding `inverse' process (cyan, raw data; blue, running average) of starting from $Ra_0 = 8.40\times10^7$ with $c = -20$. (c) Reverse hysteresis loop between chaos and turbulence in the $Nu-Ra$ plane. For clarity, only running averaged forward (red) and inverse (blue) curves are shown. Arrows mark the locations where transitions occur. Open and solid symbols denote static simulation data of chaos and turbulence, respectively.
}
\label{fig:reverse_loop}
\end{figure}

\subsubsection{The reverse hysteresis loop}

The second hysteresis loop corresponds to the transitions between chaos and turbulence. These transitions are captured in quasi-static simulations with a Rayleigh number sweeping rate of $c=\pm20$, over the range $8.1\times10^7\le Ra\le8.4\times10^7$.

Figure \ref{fig:reverse_loop}(a) shows the forward process. To highlight the transition, we plot both the raw Nusselt number time series and its running average (with a window width equal to the dominant period $\tau=281$ free-fall units). In the chaotic state, the running average of the signal is nearly constant. A clear transition to intermittent turbulence occurs around $t=75,000$ free-fall units (see the red arrow), after which both the raw and averaged signals exhibit large, irregular fluctuations characteristic of turbulence. In contrast, for the reverse process [Fig. \ref{fig:reverse_loop}(b)], the system remains in the fluctuating intermittent turbulence regime for over $91,000$ free-fall units before abruptly switching back to the oscillatory chaotic state (see the blue arrow).

Figure~\ref{fig:reverse_loop}(c) presents the running-averaged $Nu-Ra$ trajectories of both processes (red for forward, blue for reverse), overlaid with the static simulation data. The forward trajectory follows the chaotic-state data points until it abruptly drops to the intermittent turbulence branch at approximately $Ra=8.25\times10^7$ (indicated by the red arrow), after which it fluctuates around the intermittent turbulence data obtained from static simulation. The reverse trajectory instead stays on the turbulence branch far below this transition point, continuing to fluctuate down to $Ra\approx8.12\times10^7$---a value at which chaos and convection coexist in static simulations. At this point, the trajectory jumps sharply upward to rejoin the upper chaotic-state branch, and afterwards becomes indistinguishable with latter.

These forward and reverse processes form a second closed hysteresis loop. Unlike the normal loop (and classical ferromagnetic hysteresis), this loop is oriented clockwise in the $Nu-Ra$ plane, we therefore designate it the `reverse hysteresis loop'.

\begin{figure}[htbp!]
  \centerline{\includegraphics[width=1\columnwidth]{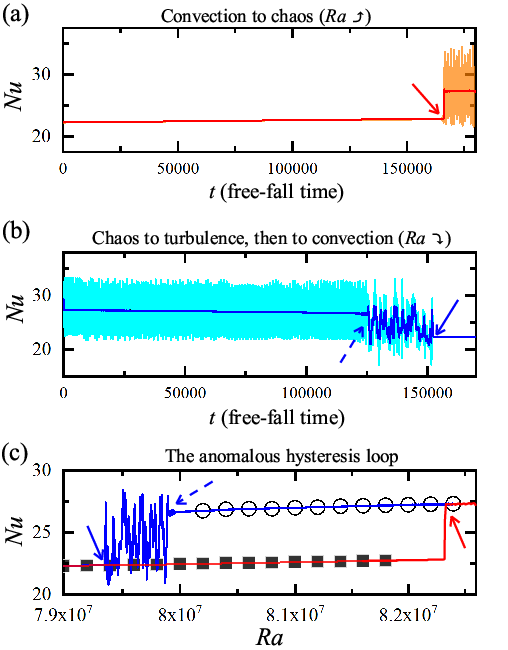}}
  \caption{(a) The time series of $Nu_t$ during a forward quasi-static process (orange, raw data; red, running average) starting from $Ra_0 = 7.9\times10^7$ with a rate of $c = 20$. (b) The time series of an inverse process (cyan, raw data; blue, running average) starting from $Ra_0 = 8.24\times10^7$ with $c = -20$. (c) The `anomalous' hysteresis loop between convection and chaos, in which intermittent turbulence serves as a transient state. For clarity, only running-averaged forward (red) and inverse (blue) curves are plotted, with arrows marking the locations where transitions occur. The open circles and solid squares denote static simulation data of chaos and convection, respectively.}
\label{fig:anomalous_loop}
\end{figure}

\subsubsection{The anomalous hysteresis loop}
The third hysteresis loop corresponds to the mutual transitions between steady convection and oscillatory chaos observed in our static simulations. It is captured in quasi-static simulations with $c=\pm20$ over the range $7.9\times10^7\le Ra\le8.24\times10^7$, as shown in Fig.~\ref{fig:anomalous_loop}.

Figure~\ref{fig:anomalous_loop}(a) presents the forward process, starting from the convective state at $Ra_0=7.9\times10^7$. The system remains in steady convection, with a linearly increasing $Nu$, for about $166,000$ free-fall units, before abruptly jumping to the high-$Nu$ chaotic state (see the red arrow). The reverse process [Fig.~\ref{fig:anomalous_loop}(b)] is more complex. Starting from the chaotic state at $Ra_0=8.24\times10^7$, the system remains chaotic for about $125,000$ free-fall units. It then transitions into a meta-stable intermittent turbulent state with large, random fluctuations in $Nu$ (see the dashed arrow), persisting for another $28,000$ time units before finally relaminarizing into the convective state (see the solid blue arrow). This long-lived transient intermittent turbulence state is also consistent with the behavior seen in static simulations [Fig.~\ref{fig:transitions}(b)].

The combined forward and reverse trajectories form a third closed hysteresis loop [Fig. \ref{fig:anomalous_loop}(c)]. Unlike the previous two loops, this one features a transient turbulence state that appears only during the transition between convection and chaos. We therefore designate it as the `anomalous hysteresis loop'.

The normal loop describes the direct transition between convection and turbulence. The reverse loop is oriented counterclockwise, representing the transitions between chaos and turbulence. The anomalous loop between convection and chaos involves a transient, meta-stable turbulence state. These three coexisting hysteresis loops collectively demonstrate the multistability, path-dependent and irreversible character of the flow transitions, suggesting a strong subcritical nature of the laminar-to-turbulent transition in quasi‑1D RBC.

\begin{figure*}[htbp!]
  \centerline{\includegraphics[width=2\columnwidth]{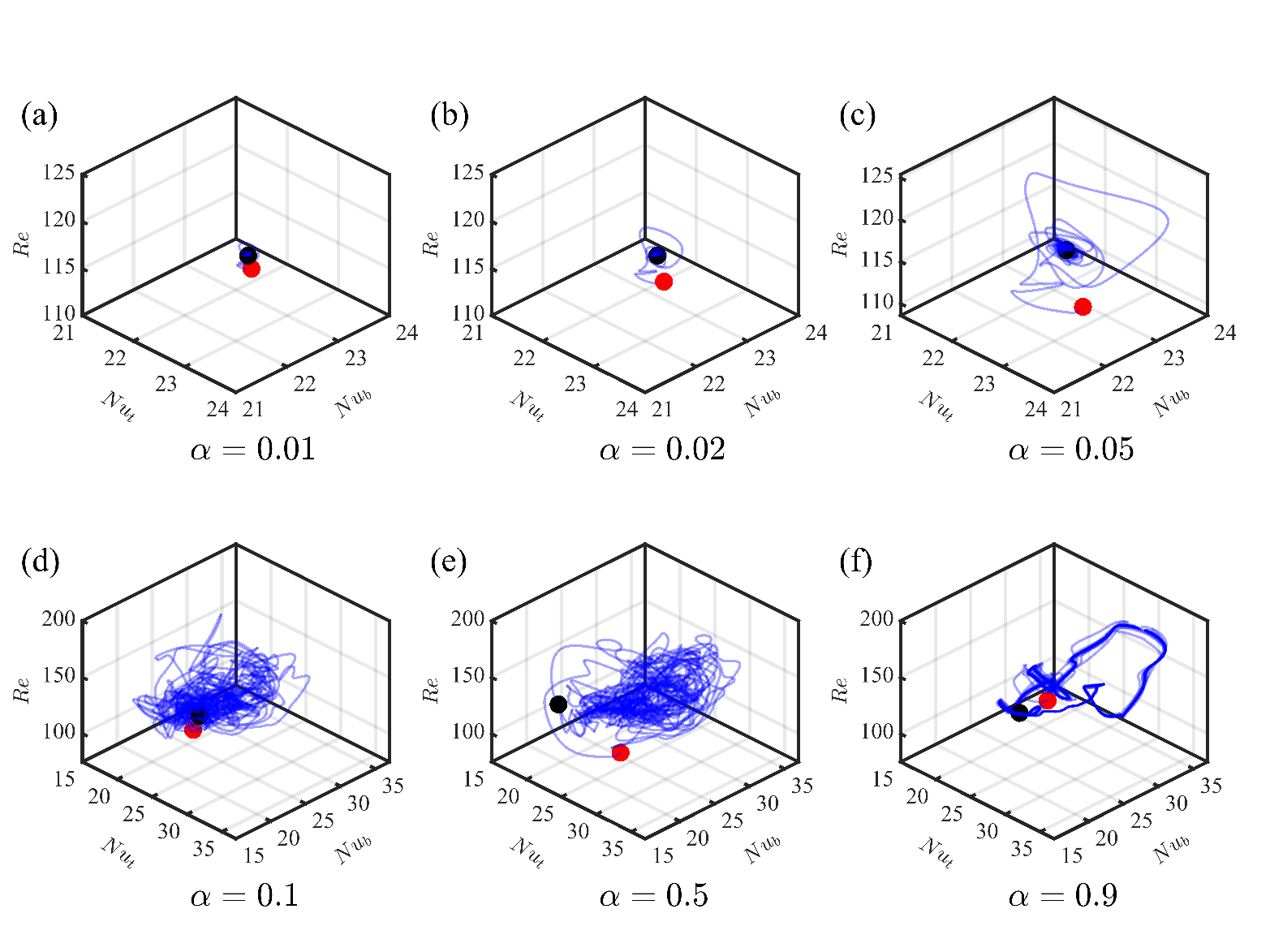}}
  \caption{Evolution trajectories of the quasi-1D RBC system, projected onto the $Nu_t$–$Nu_b$–$Re$ phase space, under initial conditions with six different perturbation amplitudes \(\alpha\) (panels (a)–(f)). The red and black dots denote the starting and ending points of each trajectory.}
\label{fig:stability}
\end{figure*}

\subsection{Stability of the steady convection state}

A defining feature of subcritical laminar-to-turbulent transition is that the laminar state is linearly stable against infinitesimal perturbations, yet susceptible to finite-amplitude disturbances \cite{Pringle_PRL_2010,Rabin_2012_JFM}. To directly verify whether the steady convective state in quasi-1D RBC exhibits this characteristic stability, we performed a series of perturbation-response simulations.

We take the steady convective state at $Ra=8.10\times10^7$ as the base flow ${U,T}$. A snapshot taken from the coexisting chaotic state at the same $Ra$, denoted as ${U^\prime,T^\prime}$, is used to construct finite-amplitude perturbations of varying strength: $U(\alpha)=U+\alpha(U^\prime-U), T(\alpha)=T+\alpha(T^\prime-T)$, where $\alpha$ represents the amplitude of perturbation. Each perturbed state ${U(\alpha),T(\alpha)}$ is assigned as the initial fields and is evolved for $5\times10^3$ free-fall time units. Figure \ref{fig:stability} shows the resulting trajectories projected onto the three-dimensional phase space spanned by the top-plate Nusselt number $Nu_t$, the bottom-plate Nusselt number $Nu_b$, and the volume-averaged Reynolds number $Re$. 

The behavior of the trajectories reveals a clear amplitude-dependent stability. For small perturbation amplitudes $\alpha\le0.05$, [Figs.~\ref{fig:stability}(a-c)], the system converges to the fixed-point attractor, confirming that the steady flow is stable to weak perturbations. However, when the perturbation amplitude exceeds a finite threshold $\alpha\ge0.1$, the flow no longer relaxes back to the steady convective state. Instead, it evolves towards an extended broadband attractor, signifying an intermittent turbulent state [Figs.~\ref{fig:stability}(d-e)], or the oscillatory chaotic state, manifesting as a quasi-periodic orbit in the phase space projection (Fig.~\ref{fig:stability}(f)).  

The results above provide direct evidence for the subcritical nature of the convection-to-turbulence transition in quasi‑1D RBC. The coexistence of linear stability against infinitesimal perturbations and finite-amplitude instability is a hallmark of subcritical transitions, which is well-established in shear-driven flows such as pipe flow but has not previously been demonstrated in buoyancy-driven convection systems.

\section{Conclusions and outlooks\label{Conclusion}}

In this work, we study the laminar-to-turbulent transition in quasi-1D RBC, using both \textit{static} and \textit{quasi-static} numerical simulation protocols. Our primary finding is the discovery of a clear subcritical transition to turbulence in a buoyancy-driven flow---a route previously thought to be exclusive to shear-driven systems. 

We provide several evidences for the subcritical transition. Firstly, we observe three coexisting flow states---steady convection, oscillatory chaos, and intermittent turbulence---over a narrow $Ra$ range, with pronounced differences in their global heat and momentum transport efficiencies, respectively. Secondly, quasi-static simulations reveal three distinct hysteresis loops, each governing the transitions between two specific flow states, which also confirms the irreversible and path-dependent nature of the transitions. Lastly, by imposing numerical perturbations, we show that the steady convection is linearly stable but loses stability to finite-amplitude perturbations, satisfying the essential characteristics of subcriticality.

The present findings challenge the long-held view that Rayleigh--B\'enard convection only exhibits supercritical `laminar' (or convection) to turbulent transitions. By demonstrating that geometrical confinement can fundamentally alter the nature of the transition, our results reveal the important role of dimensionality in determining the stability landscape of buoyancy-driven flows. 

The observed subcritical behaviour in quasi-1D RBC, together with long-lived meta-stable transients, and path-dependent memory effects, aligns closely with the theoretical framework developed for subcritical transitions in pipe and channel flows, thereby suggesting a possible unifying perspective across the seemingly disparate classes of buoyancy-driven and shear-driven flows.

We emphasize that the quasi-static simulation scheme introduced in this work is crucial for the discovery of three hysteresis loops---especially the normal loop---which would have been practically impossible to detect with static scheme alone. This approach provides an efficient and powerful tool for probing hysteresis in parameter regimes where conventional discrete static simulations would be prohibitively expensive or inadvertently overlook novel phenomena. The methodology is also promising for exploring new physics in a broad range of turbulent systems, from magnetohydrodynamic flows to geophysical and astrophysical flows.

Several fundamental questions remain open and call for further investigation. A primary challenge lies in developing a theoretical framework that explains the emergence of multiple hysteresis loops under quais-1D confinement from the underlying nonlinear dynamics. 
Other control parameters of the problem, for example, the aspect ratio $\Gamma$, the Prandtl number $Pr$, and the cross-sectional geometry, remain largely unexplored. Systematic investigations over a larger parameter space will be essential in determining the robustness of the observed phenomena and to map the boundaries of the subcritical regime. 

Finally, the experimental realization of the transition from convection to turbulence in quasi-1D RBC would provide indispensable support to the discoveries of the present study. Given the long timescales involved, such experiments would require careful design and meticulous measurements, but they would offer direct `real-world' evidence of the multiple hysteresis loops and subcritical transitions predicted here.
\linebreak

\appendix*
\section{The Rayleigh number variation rate in quasi-static simulation scheme}

\begin{figure}[htbp!]
	\centerline{\includegraphics[width=1\columnwidth]{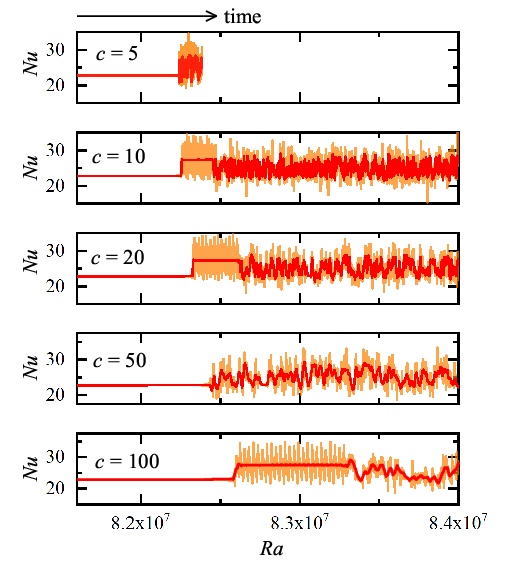}}
	\caption{The \(Nu-Ra\) curves of quasi-static simulations with different Rayleigh number increasing rate. The initial condition is a convection state at \(Ra=8.16\times10^7\) and the variation rate ranges from \(c=5\) to \(c=100\). The orange curve in each panel shows the raw data and the red curve corresponds to the running-averaged data.}
	\label{fig:ConvIncrRa}
\end{figure}

From the three hysteresis loops presented in Sec. \ref{hysteresisLoops}, we note that when the system is initially in the steady convection state and the Rayleigh number is gradually increased, the flow can transition either directly to turbulence or via an intermediate oscillatory chaotic state. The selection of the specific transition pathway appears to depend on the variation rate of Rayleigh number \(c\). We therefore perform a series of additional quasi-static simulations. The simulations are initialized from a steady convection state at \(Ra_0 = 8.16\times10^7\), and the variation rate of Rayleigh number ranges from \(c=5\) to \(c=100\). The resulting \(Nu-Ra\) curves are shown in Fig. \ref{fig:ConvIncrRa}. It is seen that the transition from the convection state appears to be random: For \(c=5\) and \(c=50\), convection directly transitions to turbulence; whereas for the other cases, the flow first bifurcates into the oscillatory chaotic state before finally transitioning to turbulence at higher $Ra$. Furthermore, we observe that as the variation rate \(c\) decreases, the bifurcation point for the transition from convection to chaos/turbulence converges towards a value of approximately \(8.22\times10^7\).

\begin{figure}[htbp!]
	\centerline{\includegraphics[width=1\columnwidth]{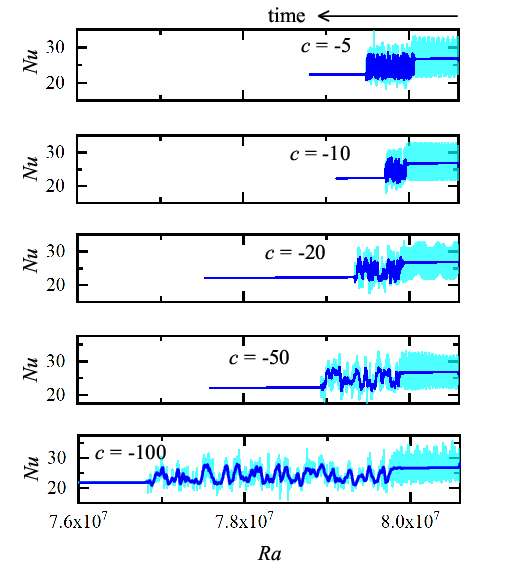}}
	\caption{The \(Nu-Ra\) curves of quasi-static simulations with different Rayleigh number decreasing rate. The initial Rayleigh number is fixed at \(Ra=8.06\times10^7\) and the variation rate ranges from \(c=5\) to \(c=100\). The orange curve in each panel shows the raw data and the red curve corresponds to the running-averaged date.}
	\label{fig:ChaoDecrRa}
\end{figure}

Additionally, we perform a set of simulations starting from the oscillatory chaotic state at \(Ra = 8.06\times10^7\), with the Rayleigh number gradually decreased at rates ranging from \(c=-5\) to \(c=-100\). The corresponding \(Nu-Ra\) curves are presented in Fig. \ref{fig:ChaoDecrRa}. The results reveal that the transition from chaos to convection, involved in the anomalous hysteresis loop, must go though a meta-stable intermediate turbulence state. Furthermore, as the magnitude of \(|c|\) decreases, the bifurcation point from the chaotic state to this intermediate turbulent state converges towards a value near \(8.00\times10^7\), which is generally consistent with the transition observed in the static simulations.

Collectively, the results above demonstrate that although the Rayleigh number variation rate $c$ introduces stochasticity in the selection of transition pathways, it does not alter the fundamental characteristics of subcritical transition, including the abrupt jump in Nusselt number and the presence of hysteresis. Therefore, the key conclusion of this work---quasi-1D RBC can undergo a subcritical laminar-to-turbulent transition---remains robust regardless of the variation rate used in the quasi-static simulations.

\color{black}

\begin{acknowledgments}
We are grateful to Jian-Jun Tao, Heng-Dong Xi, Xin Chen, and Guang-Yu Ding for stimulating discussions. This work is supported by the National Natural Science Foundation of China (Grants No.12202173, No.12595300, No.12595303 and No.12232010).
\end{acknowledgments}


\section*{Declaration of Interests}
The authors report no conflict of interest.

\section*{Author Contributions}
L.Z. and K.Q.X. designed the research; L.Z. performed simulations; both authors analyzed data and wrote the paper. 

\section*{Author ORCIDs}
L. Zhang, https://orcid.org/0000-0003-4009-2969; K.-Q. Xia, https://orcid.org/0000-0001-5093-9014.

\bibliography{ref}

\end{document}